\documentclass[twocolumn,showpacs,preprintnumbers,amsmath,amssymb]{revtex4}

\usepackage{graphicx}
\usepackage{dcolumn}
\usepackage{bm}

\begin{document}

\title{Bottom Baryon Decays in Non-relativistic Quark Model}
\author{A. Limphirat\footnotemark[1], K. Khosonthongkee\footnotemark[1]\footnotemark[2],
C. Kobdaj\footnotemark[1], P. Suebka\footnotemark[1], Y. Yan\footnotemark[1]
\vspace*{0.4\baselineskip}}
\affiliation{
\footnotemark[1]
School of Physics, Suranaree University of Technology, \\
111 University Avenue, Nakhon Ratchasima 30000, Thailand \\
\\
\footnotemark[2]
Email: khanchai@g.sut.ac.th
\vspace*{0.4\baselineskip}}

\begin{abstract}
The reactions $\Sigma _b^*  \to \Lambda _b \pi$, $\Sigma _b \to
\Lambda _b \pi$, and $\Xi_b^*  \to \Xi_b \pi$ are studied in the
$^3P_0$ non-relativistic quark model with all the model parameters
fixed in the sector of light quarks. The theoretical predictions for the decay
widths $\Gamma_{\Sigma _b^*  \to \Lambda _b \pi}$ and
$\Gamma_{\Sigma _b  \to \Lambda _b \pi}$ are consistent with the
experimental data of the CDF Collaboration. Using as an input the
recent mass of $\Xi_b$ and the theoretical predictions mass of
$\Xi_b^{*}$, a narrow decay width about 1 MeV is predicted for the
bottom baryon $\Xi_b^*$. The work suggests that the $^3P_0$ quark
dynamics is of independence of environments where heavy quarks may
or may not be a component of baryons.
\end{abstract}
\pacs{14.20.Mr,\,12.39.Jh}

\maketitle

\section{Introduction}
The first bottom baryon $\Lambda _b $, with the \emph{udb}
configuration and a mass around 5640 MeV, was reported by UA1
Collaboration at CERN in late 1990s \cite{Albajar}. Later the
$\Lambda _b $ was confirmed by other experiments such as DELPHI
Collaboration \cite{Abreu}, ALEPH Collaboration \cite{Buskulic},
and CDF Collaboration \cite{Abe} with neutral charge and mass
between 5614 to 5668 MeV. In 2005, the mass of $\Lambda _b $ was
further measured to be 5619.7 MeV by the CDF Collaboration at
Fermilab \cite{Acosta}. Very recently five new bottom baryons,
$\Sigma_b^{(*)}$ and $\Xi_b^-$ were reported by the CDF Collaboration at
Fermilab \cite{CDF,Xi0} in proton-antiproton collisions at $\sqrt s
$=1.96 TeV.

The decay processes $\Sigma _b^*  \to \Lambda _b \pi$ and $\Sigma
_b \to \Lambda _b \pi$ have been studied by combining the chiral
dynamics and the MIT bag model \cite{Hwang}, and the theoretical
results for the decay widths of the reactions are consistent
with the experimental data. More recently, the strong decays of
$\Sigma _b^{(*)}$ and $\Xi_b^*$ are studied in the $^3P_0$ quark
model, as a byproduct of the work \cite{XiangLiu} which
concentrates on the strong decays of charmed baryons. However, the
limited consistency of the theoretical results with the experimental data
make it rather difficult to conclude whether the $^3P_0$ quark dynamics,
with all the model parameters fixed in the light quark sector, is
applicable to the sector of bottom baryons.

The observation of the four bottom baryons $\Sigma_b^{(*)\pm}$ in
the $\Lambda_b\pi$ invariant mass spectrum make it possible to
explore whether the $^3P_0$ non-relativistic quark dynamics is
independent of environments which may or may not have heavy quarks
involved. In this work we study the decay processes $\Sigma _b^* \to
\Lambda _b \pi$, $\Sigma _b \to \Lambda _b \pi$, and $\Xi_b^* \to
\Xi_b \pi$ in the $^3P_0$ quark dynamics with all the model
parameters fixed by reactions in the light quark sector. The paper
is arranged to calculate the widths of the $\Sigma _b^{(*)}$ and
$\Xi_b^*$ strong decays in Section II and to give our discussion and
conclusions in Section III.
\section{$\Sigma _b^{(*)}$ and $\Xi_b^*$ decay in the $^3P_0$ QUARK DYNAMICS}
\begin{figure}[h!]
\includegraphics[width=0.45\textwidth]{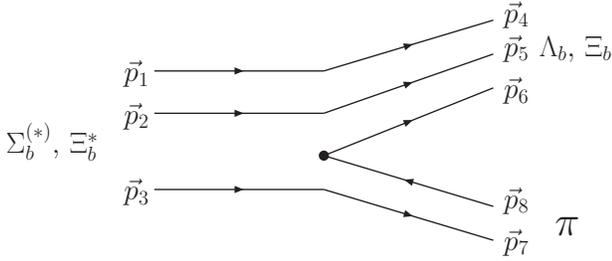}
\caption{\label{feyn1} Diagram for $\Sigma _b^{(*)}  \to \Lambda _b \pi $ and
$\Xi^*_b \to \Xi_b\pi$ in the $^3P_0$ quark model.}
\end{figure}

We study here the decay processes $\Sigma _b^{(*)} \to \Lambda _b
\pi $ and $\Xi^*_b \to \Xi_b\pi$ in the $^3P_0$ quark model. There
is no experimental data for the masses of $\Xi_b^{*}$, but one may
make a reasonable estimation by averaging the predictions of recent
theoretical works \cite{Xi1,Xi2,Xi3,Xi4,Xi5,Xi6}. The theoretical
predictions are indeed very close each other, and the averaged value
for the $\Xi_b^{*}$ mass is 5967 MeV. With such a mass, the
$\Xi_b^{*}$ may decay strongly via only one channel, the decay
process $\Xi^*_b \to \Xi_b\pi$. The transition amplitudes of the
decay processes $\Sigma _b^{(*)} \to \Lambda _b \pi $ and $\Xi^*_b
\to \Xi_b\pi$ in the $^3P_0$ quark model shown in Fig. \ref{feyn1}
are defined as
\begin{equation}\label{eqn::1}
T = \left\langle \Psi _{f} \right|V_{68}\left| \Psi _{i}
\right\rangle
\end{equation}
where $\Psi_f$ and $\Psi_i$ are respectively the final and initial
states of the reactions. $V_{68}$ is the quark-antiquark
$^3P_0$ vertex, taking the form
\begin{eqnarray}\label{vertexij}
 V_{ij} &=&
\lambda\,\vec\sigma_{ij}\cdot(\vec
p_{i}-\vec p_{j})\,\hat C_{ij}\,\hat F_{ij}\,\delta(\vec p_{i}+\vec p_{j}) \nonumber \\
&=& \lambda\sum_{\mu}\sqrt{\frac{4\pi}{3}}
(-1)^\mu\sigma^\mu_{ij}\,y_{1\mu}(\vec p_{i}-\vec p_{j})\,\hat C_{ij}\hat F_{ij}\,\delta(\vec p_{i}+\vec p_{j})
\nonumber \\
\end{eqnarray}
with
\begin{eqnarray}
\vec\sigma_{ij}=\frac{\vec\sigma_{i}+\vec\sigma_{j}}{2}
\end{eqnarray}
\begin{eqnarray}
y_{1\mu}(\vec p)\equiv |\vec p\,|\,Y_{1\mu}(\hat p)
\end{eqnarray}
where $\hat p\equiv\vec p/|\vec p\,|$, $\sigma_{i}$ are Pauli
matrices and $Y_{1m}(\hat p)$ are the spherical harmonics. $\vec
p_i$ and $\vec p_j$ are the momenta of quark and antiquark which
pumped out from vacuum, and $\hat C_{ij}$ and $\hat F_{ij}$ are
respectively the color and flavor operators projecting a
quark-antiquark pair to vacuum in the color and flavor spaces. The
derivation and interpretation of the quark-antiquark $^3P_0$
dynamics may be found in literatures \cite{3P0,tueb1}.

The spin-flavor wave functions of the baryons involved
may be constructed in the framework
of the flavor SU(4) and spin SU(2) symmetries. The explicit forms of
the spin-flavor functions are given in Appendix A. As for the
spatial wave functions which depend on the strong interaction, we
just adopt the conventional Gaussian form which result from the
spherical harmonics oscillator interaction. We have, for example,
for $\pi$ meson
\begin{equation}\label{eqn::3}
\psi(\pi ) = N_m\exp \left[ { - \frac{{b^2 }} {8}\left(
{\overset{\lower0.5em\hbox{$\smash{\scriptscriptstyle\rightharpoonup}$}}
{p} _1  -
\overset{\lower0.5em\hbox{$\smash{\scriptscriptstyle\rightharpoonup}$}}
{p} _2 } \right)^2 } \right]
\end{equation}
and for $\Sigma_b$
\begin{equation}\label{eqn::4}
\begin{gathered}
  \psi\left( {\Sigma _b } \right) = N_B\exp \left[ { - \frac{{a^2 }}
{2}\left(
{\frac{{\overset{\lower0.5em\hbox{$\smash{\scriptscriptstyle\rightharpoonup}$}}
{p} _2  -
\overset{\lower0.5em\hbox{$\smash{\scriptscriptstyle\rightharpoonup}$}}
{p} _3 }} {{\sqrt 2 }}} \right)^2 } \right] \hfill \\
  \;\;\;\;\;\;\;\;\;\;\;\;\;\;\cdot\exp \left[ { - \frac{{a^2 }}
{2}\;d^2\left(
{\frac{{\overset{\lower0.5em\hbox{$\smash{\scriptscriptstyle\rightharpoonup}$}}
{p} _2  +
\overset{\lower0.5em\hbox{$\smash{\scriptscriptstyle\rightharpoonup}$}}
{p} _3  -
2m_r\overset{\lower0.5em\hbox{$\smash{\scriptscriptstyle\rightharpoonup}$}}
{p} _1 }}
{{\sqrt 6 }}} \right)^2 } \right] \hfill \\
\end{gathered}
\end{equation}
with $N_m=b^{3/2}/\pi^{3/4}$, $N_B=3^{3/4}a^3/\pi^{3/2}$,
$m_r=m_q/m_b$ and $d=3/(1+2m_r)$ where $m_q$ and $m_b$ are
respectively the masses of the constituent $u(d)$ and $b$
quarks. The parameters $b$ and $a$ in Eqs. (\ref{eqn::3}) and (\ref{eqn::4})
are linked to the sizes of meson and baryon, respectively.

The evaluation of the transition amplitudes is straightforward for
all the decay processes, and it is found that only the $l=1$ partial wave
gives contributions. The partial wave transition amplitudes take
the general form
\begin{eqnarray}\label{eqn::5}
T_{1M}=\lambda\cdot f_1\cdot f_2\cdot f_3
\end{eqnarray}
with $f_1$, $f_2$ and $f_3$ resulting respectively from the
spin, spatial and color-flavor sectors. Detailed calculations lead to
\begin{eqnarray}\label{eqn::6}
f_1=C(S_iM_i,1M;S_fM_f)\left[
\begin{array}{ccccc}
1 & & 1/2 &&  S_i \\
1/2 && 1/2 && 1 \\
S_f && 0 && S_f
\end{array}
\right]
\end{eqnarray}
\begin{eqnarray}\label{eqn::7}
f_2=\frac{16\,\pi^5\left((9m_r+3) a^2+b^2 (2m_r+1)\right)
}{9 a^3 \left(3 a^2+b^2\right)^{5/2} (2m_r+1)}
\end{eqnarray}
where
$S_i$ and $S_f$ are respectively the spins of the initial and final baryons,
being $\frac{3}{2}$
for $\Sigma _b^{*}$ and $\Xi_b^*$, and $\frac{1}{2}$ for $\Lambda_b$,
$\Xi_b$ and
$\Sigma_b$. $M_i$ and $M_f$ are the corresponding spin magnetic moments.
The first and second factors in Eq. (\ref{eqn::6}) are respectively
the C-G coefficient
and square $9j$ symbol. The factor $f_3$ in Eq. (\ref{eqn::5}) takes the
values as
\begin{eqnarray}
f_3=\left\{
\begin{array}{ccc}
\frac{1}{3}\,, & & \Sigma _b^{*\pm} \to \Lambda _b \pi^{\pm} \\
\\
\frac{1}{3\sqrt{2}}\,, & & \Sigma _b^{\pm} \to \Lambda _b \pi^{\pm} \\
\\
\frac{1}{3\sqrt{2}}\,, & & \Xi^{*-}_b \to \Xi_b^{0}\pi^- \\
\\
\frac{1}{6}\,, & & \Xi^{*-}_b \to \Xi_b^{-}\pi^0
\end{array}
\right.
\end{eqnarray}
The decay width of the processes $\Sigma _b^{(*)} \to \Lambda _b
\pi $ and $\Xi^*_b \to \Xi_b\pi $ takes the form in terms of the partial wave
transition
amplitudes \cite{Joachain}
\begin{eqnarray}
\Gamma= \frac{2\pi E_1 E_2 k}{M_B}
\,\frac{1}{2S_i+1}\sum_{M,M_i}\left|T_{1M} \right|^2
\end{eqnarray}
where $k$ is the final momentum at the rest frame of the initial particle, $M_B$ the
mass of the initial baryon, and
$E_1$ and $E_2$ are the energies of the two final particles.

In addition to the quark masses, one also needs to determine, prior to
our evaluation of the decay widths of $\Sigma^{(*)}$ and $\Xi^*$,
the effective strength parameter
$\lambda$ of
the $^3P_0$ quark vertex and the baryon and meson size parameters
$a$ and $b$. We take for the $u$ and $d$ quarks
the widely used constituent quark mass $m_u=m_d=330$ MeV, and for the $s$
quark $m_s=550$ MeV. For the $b$ quark we use the $\overline{\mbox MS}$
mass $m_b=4.2$ GeV evaluated by the Particle Data Group \cite{PDG}.
The
meson size parameter $b$ in the work is determined to be 3.85 GeV$^{-1}$
by the reaction
$\rho \to e^+e^-$ as in the work \cite{Yan} while the value of the baryon size
parameter $a$ is taken to be 3.1 GeV$^{-1}$ which corresponds to a 0.6 fm
quark core of ground state baryons \cite{tueb1,asize}.
\begin{table}
\begin{center}
\label{rhoomega} \caption{Summary of input parameters which are
fixed by other processes} \vspace*{.3cm}
\begin{tabular}{llll}
\hline
\hline
\\
&$\lambda$ && 3.1  \\
&$a$ && $3.1$ GeV$^{-1}$ \\
&$b$ && $3.85$ GeV$^{-1}$ \\
&$m_{u(d)}$ && 330 MeV  \\
&$m_s$ && 550 MeV  \\
&$m_b$ && 4200 MeV  \\
&$M_{\Lambda_b}$ && 5619 MeV  \\
&$M_{\Sigma_b^-}$ && 5816 MeV  \\
&$M_{\Sigma_b^+}$ && 5808 MeV  \\
&$M_{\Sigma_b^{*-}}$ && 5837 MeV  \\
&$M_{\Sigma_b^{*+}}$ && 5829 MeV  \\
&$M_{\Xi_b}$ && 5793 MeV  \\
&$M_{\Xi_b^{*}}$ && 5967 MeV  \\
\\
\hline
\hline
\end{tabular}
\end{center}
\end{table}
\begin{table}[h]
\begin{center}
\label{rhoomega} \caption{Decay widths (MeV) of $\Sigma_b^{(*)}
\to \Lambda_b\pi$ and $\Xi_b^{*}\to \Xi_b\pi$} \vspace*{.3cm}
\begin{tabular}{lcccc}
\hline
\hline
\\
& Reactions & $^3P_0$ results && Data \\
\\
\hline
\\
&$\Sigma_b^{*-} \to \Lambda_b\pi^-$ & 14.6  && $\sim 15$ \\
\\
&$\Sigma_b^{*+}  \to \Lambda_b\pi^+$  & 12.4  && $\sim 15$ \\
\\
&$\Sigma_b^{-} \to \Lambda_b\pi^-$  & 9.0  && $\sim 8$ \\
\\
&$\Sigma_b^{+} \to \Lambda_b\pi^+$  & 7.1 && $\sim 8$ \\
\\
&$\Xi_b^{*}\to \Xi_b\pi$ & 1.3  && $-$ \\
\\
\hline
\hline
\end{tabular}
\end{center}
\end{table}

As the main purpose
of the work is to figure out whether the $^3P_0$ quark dynamics is consistently
applicable to both the light and heavy quark sectors, we would determine the
effective coupling constant $\lambda$ via the process
$\Sigma(1385) \to \Lambda(1116)\pi$. Using as an input $b=3.85$ GeV$^{-1}$,
$a=3.1$ GeV$^{-1}$, $M_{\Sigma^+}=1383$ MeV, $M_{\Lambda}=1116$ MeV,
$m_{u}=330$ MeV, $m_{s}=550$ MeV, and
the experimental value
$\Gamma_{\Sigma^+ \to \Lambda\pi^+}=32.0$ MeV, we get the effective coupling
constant $\lambda=3.1$.

Summarized in Table I are all the input parameters for the evaluation of the
decay widths of the processes $\Sigma _b^{(*)} \to \Lambda _b
\pi $ and $\Xi^*_b \to \Xi_b\pi $. Note that all the parameters are taken from
other works.
Using as an input the parameters listed in Table I, the decay widths for
the reactions $\Sigma _b^{(*)} \to \Lambda _b
\pi $ and $\Xi^*_b \to \Xi_b\pi $ are worked out as shown
in Table II.
\section{DISCUSSION AND CONCLUSIONS}
The reactions $\Sigma _b^*  \to \Lambda _b \pi$, $\Sigma _b \to
\Lambda _b \pi$, and $\Xi_b^*  \to \Xi_b \pi$ are investigated in the
$^3P_0$ non-relativistic quark model with all the model parameters taken
from other sources.
The meson size parameter $b$ is fixed by the
reaction $\rho(770) \to e^+e^-$ while the baryon size parameter $a$ is
taken to give a 0.6 fm
quark core of ground state baryons. With $b=3.85$ GeV$^{-1}$ and $a=3.1$
GeV$^{-1}$, the effective strength parameter $\lambda$ of the $^3P_0$ quark
vertex is fixed by
the reaction $\Sigma(1385) \to \Lambda(1116)\,\pi$. The theoretical
prediction for the decay width of the process $\Sigma(1385) \to \Lambda(1116)\,\pi$
depends strongly on the size parameters $a$ and $b$, hence with different values
of $a$ and $b$ we surely needs different $\lambda$ to fit the experimental
decay width. However, the predictions for the decay widths of
the decay processes $\Sigma _b^*  \to \Lambda _b \pi$, $\Sigma _b \to
\Lambda _b \pi$, and $\Xi_b^*  \to \Xi_b \pi$ are rather insensitive to
the combined parameter set $\{a,\, b,\, \lambda\}$.
For instance, a 20\% variation of
$a$ and $b$ results in less than 5\% change over the decay widths.

The theoretical predictions for the decay widths of
the decay processes $\Sigma _b^*  \to \Lambda _b \pi$, $\Sigma _b \to
\Lambda _b \pi$, and $\Xi_b^*  \to \Xi_b \pi$ are also insensitive to the quark masses. With
different masses of $u(d)$ and $s$ quarks one gets different coupling
constants $\lambda$ from fitting to the experimental decay width of the
process $\Sigma(1385) \to \Lambda(1116)\pi$, but the combined effect
is very trivial on the theoretical predictions for the decay widths of the reactions
$\Sigma _b^*  \to \Lambda _b \pi$, $\Sigma _b \to
\Lambda _b \pi$, and $\Xi_b^*  \to \Xi_b \pi$. Since the mass of the bottom quark
is much larger than the light ones, a 10\% variation of the mass $m_b$ about 4.2 GeV
gives no observable effect on the theoretical predictions.

The predictions for the decay widths of the reactions
$\Sigma _b^*  \to \Lambda _b \pi$ and $\Sigma _b \to
\Lambda _b \pi$ are in line with the CDF experimental data \cite{CDF}.
One may conclude that the $^3P_0$ quark dynamics is
of independence of environments where heavy quarks may or may not be
a component of baryons.

Using as an input $M_{\Xi_b}=5793$ MeV from the experimental data
and $M_{\Xi_b^{*}}=5967$ MeV derived by averaging the recent
theoretical predictions, the work predicts a narrow $\Xi_b^*$ width
$\Gamma\approx 1$ MeV.

\section*{ACKNOWLEDGEMENTS}

This work is supported in part by the National Research Council of
Thailand (NRCT) under Grant No. 1.CH5/2549 and Commission on Higher
Education, Thailand (CHE-RES-RG "Theoretical Physics").

\appendix
\section{SPIN-FLAVOR WAVE FUNCTIONS}
The spin-flavor
wave functions of baryons made of $u$, $d$, $s$ and $b$ quarks
may be constructed in the framework of the flavor
SU(4) and spin SU(2) symmetries. The spin-flavor wave functions
$\Psi_{SF}$ for the baryons $\Sigma _b^{*+}(uub)$, $\Sigma _b^+
(uub)$, $\Xi_b^{*0}(usb)$, $\Xi_b^{0}(usb)$,
 and $\Lambda_b^0(usb)$ are respectively
\begin{equation}
\begin{gathered}
  \Psi _{SF} (\Sigma _b^{* + } )  = \phi^S(\Sigma _b^{* + } )\,\chi ^S \hfill \\
   \Psi _{SF} (\Xi_b^ {*0})  =  \phi^S(\Xi_b^ {*0})\,\chi ^S \hfill \\
    \Psi _{SF} (\Xi_b^0 )
  = \frac{1}{\sqrt{2}}\left[ {\phi ^\lambda(\Xi_b^0 )\,\chi ^\lambda
  + \phi ^\rho(\Xi_b^0 )\,\chi ^\rho  } \right] \hfill \\
   \Psi _{SF} (\Sigma _b^ +  )
  = \frac{1}{\sqrt{2}}\left[ {\phi ^\lambda(\Sigma _b^ +  )\,\chi ^\lambda
  + \phi ^\rho(\Sigma _b^ +  )\,\chi ^\rho  } \right] \hfill \\
     \Psi _{SF} (\Lambda_b^0 )
  = \frac{1}{\sqrt{2}}\left[ {\phi ^\lambda(\Lambda_b^0 )\,\chi ^\lambda
  + \phi ^\rho(\Lambda_b^0 )\,\chi ^\rho  } \right] \hfill \\
\end{gathered}
\end{equation}
where $\chi^S$ ($\phi^S$), $\chi^\lambda$ ($\phi^\lambda$)
and $\chi^\rho$ ($\phi^\rho$) are the symmetric, $\lambda$ type and
$\rho$ type spin (flavor) wave functions, respectively.

It is convenient to construct wave functions of baryons or other
multi-quark particles in the framework of the Yamanouchi basis
and the corresponding projection operators of permutation group.
For more details, one may refer to group theory books like \cite{Yan1,Chen}.
The various flavor wave functions are
\begin{equation}
\begin{gathered}
  \phi ^S \left(\Sigma _b^{* + } \right) = \frac{1}
{\sqrt 3 }\left( {uub + buu + ubu} \right) \hfill \\
  \phi ^S \left(\Xi_b^ {*0}\right) = \frac{1}
{\sqrt 6 }\left( {usb + bsu + sbu + sub+bus+ubs} \right) \hfill \\
  \phi ^\lambda  (\Sigma _b^ +  ) = \frac{1}
{\sqrt 6 }\left( {2uub - ubu - buu} \right) \hfill \\
  \phi ^\rho  (\Sigma _b^ +  ) = \frac{1}
{\sqrt 2 }\left( {buu - ubu} \right) \hfill \\
  \phi ^\lambda  (\Xi_b^0 ) = \frac{1}
{2}\left( {ubs+bus-bsu-sbu} \right) \hfill \\
  \phi ^\rho  (\Xi_b^0 ) = \frac{1}
{\sqrt {12} }\left( {2sub-2usb+bus+sbu-ubs-bsu} \right) \hfill \\
  \phi ^\lambda  (\Lambda_b^0 ) = \frac{1}
{2}\left( {ubd+bud-bdu-dbu} \right) \hfill \\
  \phi ^\rho  (\Lambda_b^0 ) = \frac{1}
{\sqrt {12} }\left( {2dub-2udb+bud+dbu-ubd-bdu} \right) \hfill \\
\end{gathered}
\end{equation}
where $u$, $d$, $s$ and $b$ stand for the flavor
wave functions of the corresponding quarks, respectively.

\end{document}